\documentclass[12pt,a4paper]{ouparticle}
\usepackage{cite}
\usepackage{graphicx}
\usepackage{adjustbox}
\usepackage{xcolor}
\usepackage{float}
\usepackage[utf8]{inputenc}
\usepackage{tikz}
\usepackage{booktabs} 
\usetikzlibrary{shapes.geometric, arrows.meta, positioning}
\usepackage{longtable}
\usetikzlibrary{shapes.geometric, arrows.meta}
\usepackage[colorlinks=true, citecolor=blue]{hyperref}
\begin{document}

\title{A comparative study of sum-connectivity and product-connectivity Gourava indices for benzenoid hydrocarbons}
\author{%
\name{H. M. Nagesh}
\address{Department of Science and Humanities,\\ PES University,
Bangalore, Karnataka, India.}
\email{hmnagesh1982@gmail.com}
\name{U. Vijaya Chandra Kumar}
\address{Department of Mathematics, School of Applied Sciences,\\
REVA University, Bengaluru, Karnataka, India.}
\email{upparivijay@gmail.com}
\name{Azghar Pasha B}
\address{Department of Mathematics, \\
M. S. Ramaiah Institute of Technology,\\ Bengaluru, Karnataka, India.}
\email{azgharpasha@msrit.edu}
\name{Narahari. N}
\address{Department of Mathematics,\\ 
University College of Science, \\
Tumkur University, Tumakuru, Karnataka, India.}
\email{narahari\_nittur@yahoo.com}
}

\abstract{This study evaluates the sum-connectivity ($SGO$) and product-connectivity ($PGO$) Gourava indices as molecular descriptors for benzenoid hydrocarbons. Using a dataset of 30 benzenoid structures, we compare least-squares regression models for predicting $\pi$-electronic energies ($E_{\pi}$) and find that $SGO$ yields a markedly better fit than $PGO$ across molecular edge types. The indices are further assessed using three validation designs: (i) correlation analysis, in which $SGO$ exhibits strong yet non-perfect inverse correlations with standard descriptors ($M_1, M_2, SO, DSO,$ and $ABS$; $r\in[-0.9923,-0.8936]$), suggesting complementary structural information; (ii) degeneracy analysis on Octane, Nonane, and order-$10$ tree datasets, where $SGO$ attains low degeneracy rates (22.22\%, 40.00\%, and 42.45\%); and (iii) structure-sensitivity analysis on trees of order $n=10$, showing 74\% higher sensitivity than $DSO$ while maintaining a high structure-abruptness ratio ($SA = 0.474386$). Overall, $SGO$ offers a favorable balance between discriminative power and numerical stability, supporting its applicability in QSPR modeling and related theoretical studies.
}

\date{}

\newpage

\maketitle

\section{Introduction}
\label{sec1}
In chemical graph theory, a molecular structure is represented as a graph $G = (V, E)$, where the vertices $V$ correspond to atoms and the edges $E$ correspond to chemical bonds. A topological index is a numerical descriptor derived from this graph that remains invariant under graph isomorphism. These indices are widely utilized in Quantitative Structure-Property Relationship (QSPR) and Quantitative Structure-Activity Relationship (QSAR) studies to predict various physicochemical properties of molecules, such as boiling points, stability, electronic energy, etc. Among the various classes of topological descriptors, vertex-degree-based indices have received significant attention \cite{1,2,3}. 

The sum-connectivity and product-connectivity Gourava indices were introduced by V. R. Kulli \cite{4,5} as modern additions to the framework of vertex-degree-based molecular descriptors. These indices provide a refined approach to capturing molecular branching and connectivity by combining the sums and products of vertex degrees \cite{6}. Let $d(u)$ and $d(v)$ denote the degrees of the vertices $u$ and $v$ that are connected by an edge $uv$. The \emph{sum-connectivity Gourava index}, denoted by $SGO(G)$, and the \emph{product-connectivity Gourava index}, denoted by $PGO(G)$, are defined as follows:

\begin{equation*}
SGO=SGO(G)= \sum_{uv \in E(G)} \frac{1}{\sqrt{d(u) + d(v) + d(u)d(v)}}
\end{equation*}

\begin{equation*}
PGO=PGO(G)= \sum_{uv \in E(G)} \frac{1}{\sqrt{(d(u) + d(v))(d(u)d(v))}}
\end{equation*}

In this study, we focus on the application of these Gourava indices to benzenoid hydrocarbons. Benzenoid systems form a fundamental class of polycyclic aromatic hydrocarbons composed of fused benzene rings. We evaluate the predictive power of these indices by correlating them with $\pi$-electronic energies, demonstrating their effectiveness as reliable topological descriptors for aromatic molecular systems. Figure 1 illustrates the different types of inlets occurring in benzenoid systems \cite{8}.

\begin{figure}[H]
    \centering
    \begingroup
    \setlength{\fboxsep}{0pt}
    \colorbox{black!15}{%
        \includegraphics[width=0.4\linewidth]{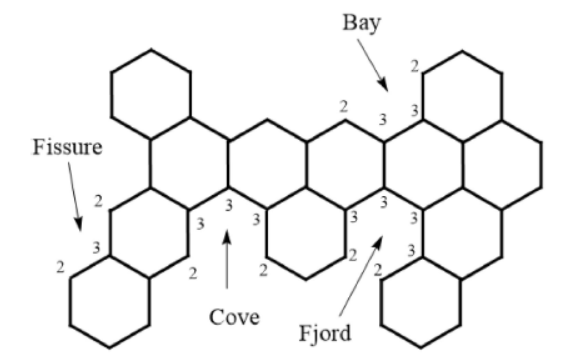}%
        }
    \endgroup
    \caption{Representation of bays, fjords, fissures, and coves in benzenoid systems.}
    \label{fig:placeholder}
\end{figure}

\subsection{Motivation} 
The motivation for this study arises from the well-established effectiveness of classical connectivity indices in characterizing polycyclic aromatic systems. Earlier studies \cite{7} have shown that the product-connectivity (Randi\'{c}) index and its variant, the sum-connectivity index, exhibit strong correlations with the $\pi$-electronic energies of benzenoid hydrocarbons. In light of the recent introduction of Gourava indices, which offer refined versions of these connectivity measures, it is important to examine whether these newer descriptors preserve or improve upon this predictive performance. Accordingly, the present work extends the comparative framework previously applied to Randi\'{c}-type indices to the broader class of Gourava indices.

The objective of this research is to systematically evaluate the Gourava indices through a hierarchical validation process. Specifically, this study aims to:

\begin{enumerate}
    \item Comparative Prediction of $E_{\pi}$: Perform a regression analysis for $30$ benzenoid hydrocarbons to compare the predictive performance of the sum-connectivity ($SGO$) and product-connectivity ($PGO$) Gourava indices against $\pi$-electronic energies ($E_{\pi}$). 
    \item Index Selection: Demonstrate, through multilinear regression, that $SGO$ provides a statistically superior fit compared to $PGO$, thereby establishing $SGO$ as the primary descriptor for advanced topological analysis.
    \item Mathematical and Chemical Validation: Subject the superior $SGO$ index to rigorous testing—including intercorrelation analysis, degeneracy testing, and structural sensitivity (SS) analysis—across Octane, Nonane, and all trees of order $10$ datasets to validate its unique structural encoding and applicability in QSPR modeling.
\end{enumerate}

By focusing the latter half of this study on the $SGO$ index, we provide a deep characterization of the descriptor that shows the most promise for practical chemical prediction.

\section{Sum-connectivity and product-connectivity Gourava indices of benzenoid hydrocarbons}
In this section, we derive the sum-connectivity and product-connectivity Gourava indices for the benzenoid hydrocarbons depicted in Figure 2.

\begin{figure}[H]
    \centering
    \begingroup
    \setlength{\fboxsep}{0pt}
    \colorbox{black!15}{%
        \includegraphics[width=1.05\linewidth]{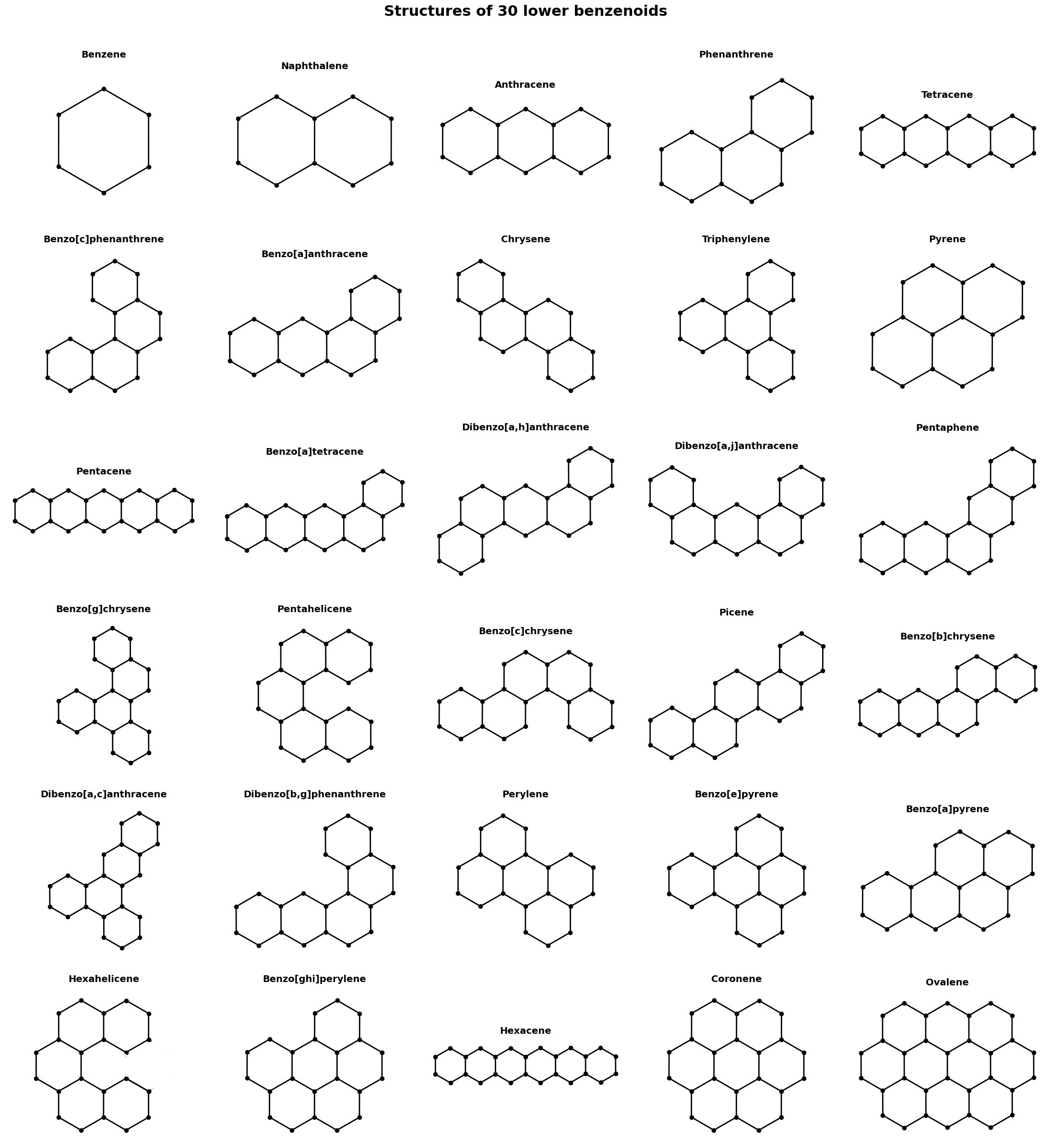}%
        }
    \endgroup
    \caption{Structures of 30 lower benzenoids.}
    \label{fig:placeholder}
\end{figure}

The structure of benzenoid hydrocarbons is most easily shown and studied using benzenoid graphs, which are mathematical drawings made of hexagonal rings. In benzenoid graphs, every vertex has a degree of either $2$ or $3$ \cite{9}, resulting in only three types of edges: $e_{22}$, $e_{23}$, and $e_{33}$. By introducing the number of these specific edges into the general definitions for Gourava indices, we obtain simplified formulas for this class of molecules. For the sum-connectivity Gourava index, we apply the edge-weights $1/\sqrt{8} \approx 0.3536$, $1/\sqrt{11} \approx 0.3015$, and $1/\sqrt{15} \approx 0.2582$ to reach the expression for $SGO(G)$. Analogously, for the product-connectivity Gourava index, the corresponding edge-weights are $1/4 = 0.25$, $1/\sqrt{30} \approx 0.1826$, and $1/\sqrt{54} \approx 0.1361$, which yield the formula for $PGO(G)$. Hence, the Gourava indices for this class of molecules are expressed as:
\begin{equation}
SGO(G) =0.3536(e_{22}) + 0.3015(e_{23}) + 0.2582(e_{33})
\end{equation}

and

\begin{equation}
PGO(G) = 0.2500(e_{22}) + 0.1826(e_{23}) + 0.1361(e_{33})
\end{equation}

Alternative expressions for (1) and (2) can be derived by considering the basic structural parameters of the graph. If a benzenoid system consists of $n$ vertices, $r$ inlets, and $h$ hexagons, the number of edges for each type is calculated as follows \cite{10}:

\begin{equation}
e_{22} = n - 2h - r + 2
\end{equation}
\begin{equation}
e_{23} = 2r
\end{equation}
\begin{equation}
e_{33} = 3h - r - 3
\end{equation}

The perimeter of these systems exhibits specific arrangements of vertex degrees called $inlets$, as established in the literature \cite{9,10,11}. These structural features are categorized into four distinct types: fissures ($f$), bays ($b$), coves ($c$), and fjords ($fj$). Each type is defined by its characteristic degree sequence along the boundary—specifically 232 for fissures, 2332 for bays, 23332 for coves, and 233332 for fjords (as shown in Figure 1). The total number of inlets, $r$, is calculated as the sum $r = f + b + c + fj$.

Substituting the edge counts from (3)-(5) into the definition for the sum-connectivity and product-connectivity Gourava indices leads to a concise formula expressed solely in terms of the number of vertices, inlets, and hexagons as follows:

\begin{gather}
SGO(G) = 0.35355n + 0.06750h - 0.00873r - 0.06750 \label{eq:6} \\
PGO(G) = 0.25000n - 0.09176h - 0.02094r + 0.09176 \label{eq:7}
\end{gather}

By utilizing the relations established in (6) and (7), we computed the sum and product connectivity Gourava indices for a set of 30 benzenoid hydrocarbons. These calculated indices, along with the $\pi$-electronic energy ($E_\pi$) values obtained from \cite{7}, are compiled and presented in Table 1.

\renewcommand{\arraystretch}{1.35}
\begin{longtable}{lccc}
\caption{$\pi$-Electronic energy ($E_{\pi}$), the sum-connectivity Gourava index ($SGO(G)$) and the product-connectivity Gourava index ($PGO(G)$) of 30 benzenoid hydrocarbons.}
\tiny
\label{tab:benzenoids} \\
\toprule
\textbf{Benzenoid Hydrocarbon} & \boldmath{$E_\pi$} & \textbf{SGO(G)} & \textbf{PGO(G)} \\
\midrule
\endfirsthead

\toprule
\textbf{Hydrocarbon Hydrocarbon} & \boldmath{$E_\pi$} & \textbf{SGO(G)} & \textbf{PGO(G)} \\
\midrule
\endhead

\bottomrule
\endfoot

Benzene & 8.0000 & 2.12130 & 1.50000 \\
Naphthalene & 13.6832 & 3.58554 & 2.36636 \\
Anthracene & 19.3137 & 5.04978 & 3.23272 \\
Phenanthrene & 19.4483 & 5.05851 & 3.25366 \\
Tetracene & 24.9308 & 6.51402 & 4.09908 \\
Benzo[c]phenanthrene & 25.1875 & 6.53148 & 4.14096 \\
Benzo[a]anthracene & 25.1012 & 6.52275 & 4.12002 \\
Chrysene & 25.1922 & 6.53148 & 4.14096 \\
Triphenylene & 25.2745 & 6.54021 & 4.16190 \\
Pyrene & 22.5055 & 5.82438 & 3.64096 \\
Pentacene & 30.5440 & 7.97826 & 4.96544 \\
Benzo[a]tetracene & 30.7255 & 7.98699 & 4.98638 \\
Dibenzo[a,h]anthracene & 30.8805 & 7.99572 & 5.00732 \\
Dibenzo[a,j]anthracene & 30.8795 & 7.99572 & 5.00732 \\
Pentaphene & 30.7627 & 7.98699 & 4.98638 \\
Benzo[g]chrysene & 30.9990 & 8.01318 & 5.04920 \\
Pentahelicene & 30.9362 & 8.00445 & 5.02826 \\
Benzo[c]chrysene & 30.9386 & 8.00445 & 5.02826 \\
Picene & 30.9432 & 8.00445 & 5.02826 \\
Benzo[b]chrysene & 30.8390 & 7.99572 & 5.00732 \\
Dibenzo[a,c]anthracene & 30.9418 & 8.00445 & 5.02826 \\
Dibenzo[b,g]phenanthrene & 30.8336 & 7.99572 & 5.00732 \\
Perylene & 28.2453 & 7.30608 & 4.54920 \\
Benzo[e]pyrene & 28.3361 & 7.30608 & 4.54920 \\
Benzo[a]pyrene & 28.2220 & 7.29735 & 4.52826 \\
Hexahelicene & 36.6814 & 9.47742 & 5.91556 \\
Benzo[ghi]perylene & 31.4251 & 8.07195 & 4.93650 \\
Hexacene & 36.1557 & 9.44250 & 5.83180 \\
Coronene & 34.5718 & 8.83782 & 5.32380 \\
Ovalene & 46.4974 & 11.85126 & 7.00664 \\
\end{longtable}

\newpage
\section{Comparative analysis of sum- and product-connectivity Gourava Indices}
To assess the statistical dependency between the sum-connectivity Gourava index and the product-connectivity Gourava index, both measures were determined for the set of 30 benzenoid systems (Table 1). The linear correlation between these two topological descriptors is expressed as follows:

\begin{equation}
PGO(G) = (0.5804 \pm 0.0065) SGO(G) + (0.3279 \pm 0.0487)
\end{equation}
\[N=30, \quad S = 0.0629, \quad R = 0.9983, \quad S_{cv} = 0.0741, \quad R_{cv} = 0.9976. \]

Here, $R$ and $R_{cv}$ represent the correlation coefficients for the model fit and the leave-one-out cross-validation, respectively. Furthermore, we reported the standard errors of fit ($S$) and leave-one-out cross-validation ($S_{cv}$), which were calculated using $N-2$ in the denominator. This statistical procedure and the validation methodology were followed as described in \cite{7}. The scatter plot illustrating this relationship is presented in Figure 3.  

\begin{figure}[H] 
    \centering
    \includegraphics[width=0.5\linewidth]{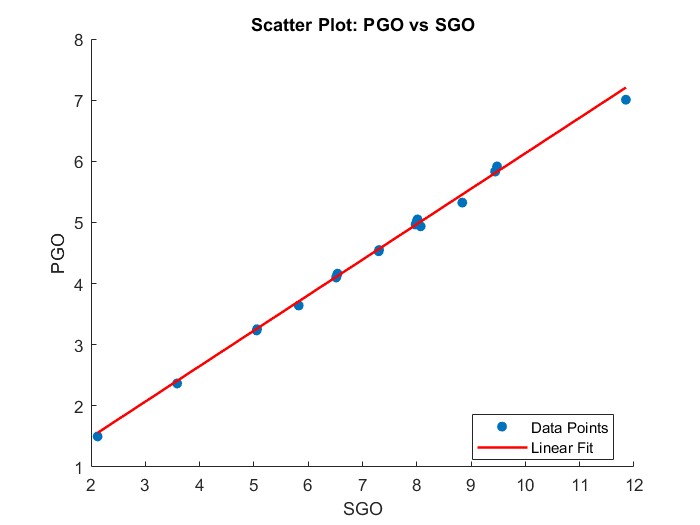} 
    \caption{Scatter plot illustrating the linear correlation between the product-connectivity Gourava index $PGO(G)$ and the sum-connectivity Gourava index $SGO(G)$ for a dataset of 30 benzenoid hydrocarbons.}
    \label{fig:scatter_plot}
\end{figure}
Both versions of the connectivity Gourava index are found to be highly intercorrelated quantities in this case ($R > 0.99$). Consequently, it is expected that the sum-connectivity Gourava index will also produce high-quality structure-property-activity relationship (QSAR/QSPR) models for benzenoid hydrocarbons, similar to its product-based counterpart. 

\section{The correlation between Gourava indices and the $\pi$-electronic energy of benzenoid hydrocarbons}

In this section, we perform a linear regression analysis between the $\pi$-electronic energy ($E_{\pi}$) and both the sum-connectivity and product-connectivity Gourava indices for the 30 lower benzenoids listed in Table 1. The resulting correlation parameters and regression equations are detailed below. 

\begin{equation}
E_{\pi} = (3.925 \pm 0.017) \, SGO(G) + (-0.460 \pm 0.132)
\end{equation}
\begin{center}
$N=30, \quad R = 0.9997, \quad S = 0.170, \quad R_{cv} = 0.9996, \quad S_{cv} = 0.194$.
\end{center}

\begin{equation}
E_{\pi} = (6.731 \pm 0.098) \, PGO(G) + (-2.538 \pm 0.460)
\end{equation}
\begin{center}
$N=30, \quad R = 0.9970, \quad S = 0.554, \quad R_{cv} = 0.9960, \quad S_{cv} = 0.644$.
\end{center}

The results presented here demonstrate that the sum-connectivity Gourava index accurately predicts the $\pi$-electronic energies of lower benzenoids, and it outperforms the product-connectivity Gourava index in terms of correlation and predictive stability. Scatter plots illustrating the relationships between $E_{\pi}$ and the sum and product-connectivity Gourava indices $SGO(G)$ and $PGO(G)$, respectively, are provided in Figure 4.

\begin{figure}[H] 
    \centering
    \includegraphics[width=0.95\linewidth]{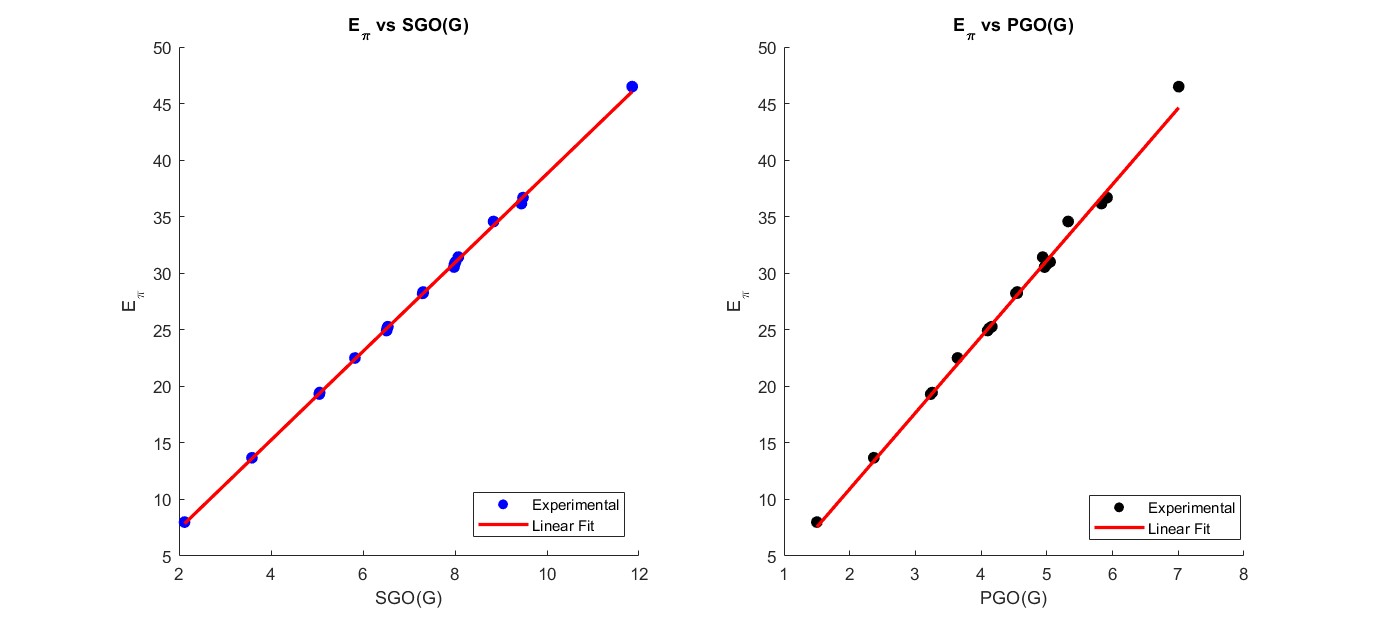} 
    \caption{Scatter plots illustrating the relationships between $E_{\pi}$, $SGO(G)$, and $PGO(G)$.}
    \label{fig:scatter_plot}
\end{figure}

As studied in \cite{7}, in addition to relationships (9) and (10), we may express $\pi$-electronic energy in the form
\begin{equation}
E_{\pi} = A+ Be_{22} + Ce_{23} + De_{33}.
\end{equation}
Regression coefficients $A,B,C$ and $D$ are obtained by the least-square fit procedure on the set of 30 benzenoid hydrocarbons from Table 1:
\begin{align}
E_{\pi} = &-0.024018(\pm 0.054239) + 1.342149(\pm 0.006416)e_{22} \nonumber \\
&+ 1.145181(\pm 0.002312)e_{23} + 1.060790(\pm 0.002566)e_{33}.
\end{align}
It is evident that the uncertainty of the intercept ($\pm 0.054$) significantly exceeds the absolute value of the parameter itself ($-0.024$). Consequently, this term is statistically insignificant and can be omitted from the model. Substituting (1) into (9) and (2) into (10) yields the following results:
\begin{equation}
E_{\pi} = -0.460(\pm 0.132) + 1.388(\pm 0.006)e_{22} + 1.183(\pm 0.005)e_{23} + 1.013(\pm 0.004)e_{33}.    
\end{equation}
\begin{equation}
   E_{\pi} = -2.538(\pm 0.460) + 1.683(\pm 0.025)e_{22} + 1.229(\pm 0.018)e_{23} + 0.916(\pm 0.013)e_{33}. 
\end{equation}

\section{Comparison Analysis}
To determine which model is better, we compare how closely the regression coefficients of the ``test" (13) and (14) match the ``optimal" coefficients in (12). This comparison is detailed in Table 2.
\begin{table}[htbp]
    \centering
    \caption{Comparison of regression coefficients for (12), (13), and (14).}
    \label{tab:coeff_comparison}
    \begin{tabular}{lcccc}
        \toprule
        Model & Intercept ($A$) & $e_{22}$ & $e_{23}$ & $e_{33}$ \\
        \midrule
        Eq. (12) Optimal & $-0.024$ & $1.342$ & $1.145$ & $1.061$ \\
        Eq. (13)         & $-0.460$ & $1.388$ & $1.183$ & $1.013$ \\
        Eq. (14)         & $-2.538$ & $1.683$ & $1.229$ & $0.916$ \\
        \bottomrule
    \end{tabular}
\end{table}

\noindent \textbf{Analysis and Interpretation:} 
One can see from the error of the regression coefficient $A$ in (12) that this parameter is statistically insignificant, as the error margin ($\pm 0.054$) exceeds the absolute value of the coefficient ($-0.024$). While (13) maintains a relatively small intercept of $-0.460$, (14) deviates significantly with a much larger negative intercept of $-2.538$. Furthermore, a comparison of the variable coefficients reveals that the values from (13) match the optimal coefficients obtained by the least-squares fit in (12) much more closely than those from (14). In every instance ($e_{22}, e_{23}$, and $e_{33}$), the parameters in (13) are nearer to the ideal values. This comparison confirms that the weighting scheme utilized in (13) provides a superior representation of the relative contribution of edge-types compared to the scheme used in (14). Consequently, the sum-connectivity Gourava index represented by (13) serves as a more accurate predictor for this specific set of molecules.


\subsection{On intercorrelations}

To evaluate the potential linear relationships among some of the prominent degree-based topological indices---specifically first Zagreb ($M_1$), second Zagreb ($M_2$), harmonic ($H$), Sombor ($SO$), diminished Sombor ($DSO$), atom-bond-sum connectivity ($ABS$), and Nirmala ($N$) ---alongside the sum-connectivity Gourava index ($SGO$), we performed a correlation analysis. The resulting correlation matrix is illustrated in Figure 5.
\newpage
\begin{figure}[hbt!] 
    \centering
    \includegraphics[width=0.70\linewidth]{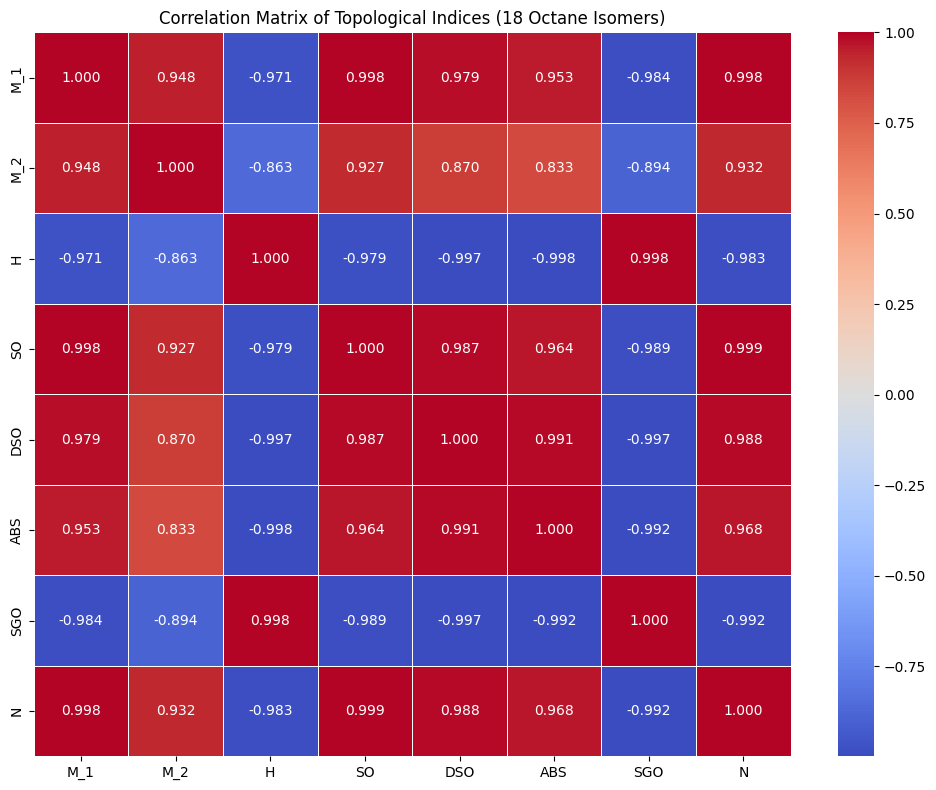} 
    \caption{Correlation matrix.}
    \label{fig:scatter_plot}
\end{figure}

The first observation is that the indices $M_1$, $M_2$, $SO$, $DSO$, $ABS$, and $N$ are highly correlated with each other. This is particularly expressed in the pair $M_1$ and $SO$, where the correlation coefficient is $0.9983$, indicating that they are very similar to each other. A slightly lower correlation is observed between the $M_2$ index and the other indices in this group, with correlation coefficients ranging from $0.8332$ to $0.9475$. 

In contrast, the $SGO$ index shows a strong negative correlation with all these indices, with coefficients ranging from $-0.8936$ to $-0.9923$, while maintaining a near-perfect positive correlation with $H$ ($0.9978$). The lowest correlation magnitude for $SGO$ is observed with the $M_2$ index ($-0.8936$). While $SGO$ shares a strong inverse relationship with the other descriptors, these values leave a noticeable gap from a perfect negative linear fit ($\vert{}r\vert{} = 1$). This variation indicates that $SGO$ is not merely a perfect inverse of the other indices; rather, it possesses a distinct level of statistical independence, particularly when compared to $M_2$. Consequently, the $SGO$ index captures unique variance and additional information regarding the graph structure that is not fully present in the other indices.

\subsection{On degeneracy} 
The degeneracy of a topological index serves as a direct measure of its capability to discriminate among non-isomorphic graphs. The property was introduced in \cite{12} as follows: 
\begin{equation*}
S_{TI}=\frac{I-I_{TI}}{I},
\end{equation*}
where $I$ is the total number of isomers/trees under consideration for testing and $I_{TI}$ is the number of them that cannot be separated by the topological index ($TI$).

Figure~6 illustrates the degeneracy analysis comparison of the sum-connectivity Gourava index ($SGO$) against seven other well-known topological indices: the first Zagreb ($M_1$), second Zagreb ($M_2$), harmonic ($H$), Sombor ($SO$), diminished Sombor ($DSO$), atom-bond-sum connectivity ($ABS$), and Nirmala ($N$) indices. \newpage The evaluation is conducted across three distinct molecular datasets: Octane ($N=18$), Nonane ($N=35$) isomers, and all trees of order 10 ($N=106$). 

\begin{figure}[hbt!]
    \centering
    \includegraphics[width=1.05\linewidth]{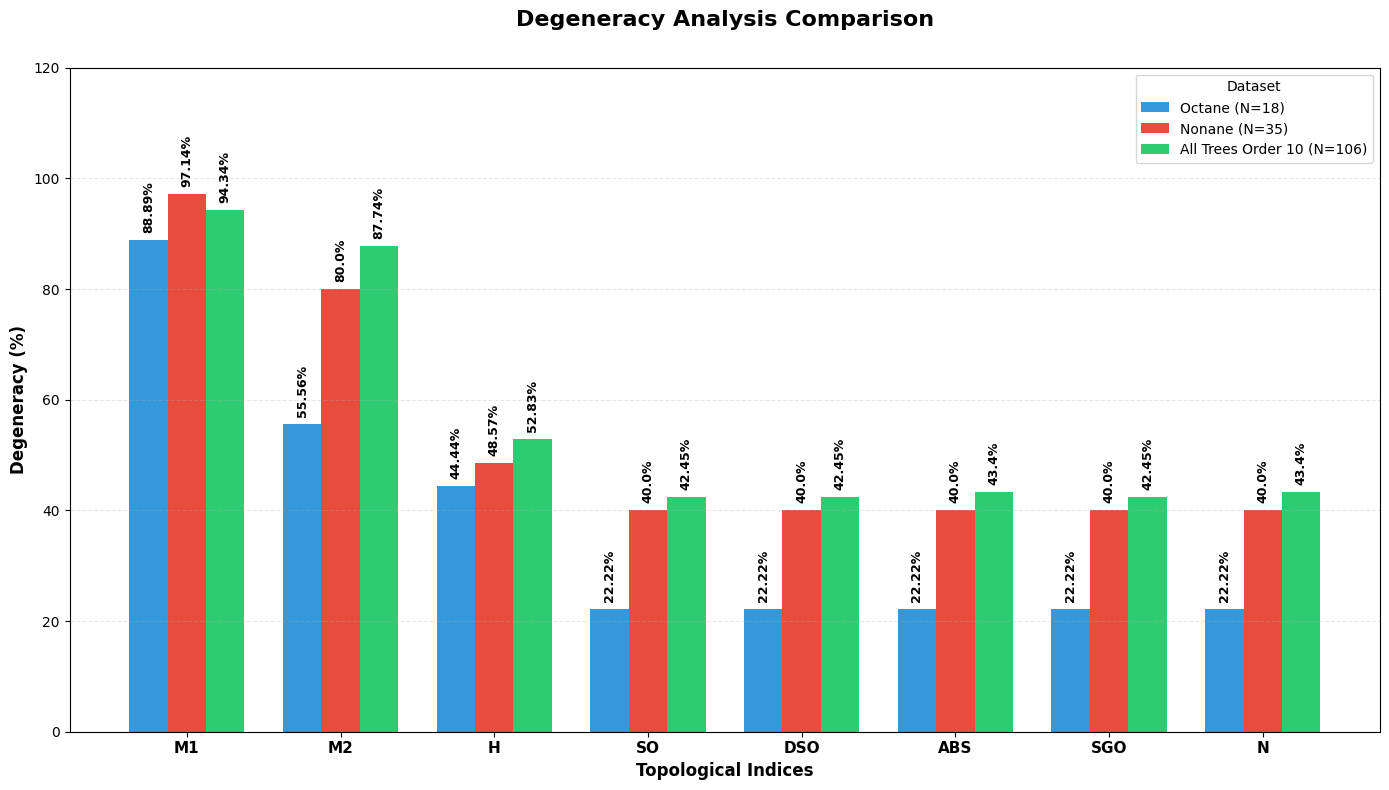}
    \caption{Correlation matrix.}
    \label{fig:scatter_plot}
\end{figure}

As established in \cite{12}, a lower degeneracy percentage indicates a higher capability of a topological index to discriminate among non-isomorphic graphs. The comparative degeneracy percentages obtained from the graphical data are organized in Table 3.

\begin{table}[htbp]
\centering
\small
\caption{Degeneracy percentage comparison among different topological indices.}
\label{tab:degeneracy_comparison}
\begin{tabular}{lcccccccc}
\toprule
\textbf{Dataset} & \textbf{$M_1$} & \textbf{$M_2$} & \textbf{$H$} & \textbf{$SO$} & \textbf{$DSO$} & \textbf{$ABS$} & \textbf{$SGO$} & \textbf{$N$} \\ 
\midrule
Octane             & 88.89\% & 55.56\% & 44.44\% & 22.22\% & 22.22\% & 22.22\% & 22.22\% & 22.22\% \\
Nonane              & 97.14\% & 80.00\% & 48.57\% & 40.00\% & 40.00\% & 40.00\% & 40.00\% & 40.00\% \\
All trees order 10  & 94.34\% & 87.74\% & 52.83\% & 42.45\% & 42.45\% & 43.40\% & 42.45\% & 43.40\% \\
\bottomrule
\end{tabular}
\end{table}

From the visual and tabulated data, the following observations can be made:
\begin{itemize}
    \item \textbf{High Degeneracy Indices:} The first Zagreb index ($M_1$) consistently exhibits the highest degeneracy across all datasets, reaching up to 97.14\% for the Nonane dataset and maintaining a high value of 94.34\% for trees of order 10. The second Zagreb ($M_2$) and harmonic ($H$) indices similarly present elevated degeneracy values, establishing their limited capability in structural isomer discrimination.
    
    \item \textbf{Performance of $SGO$:} The sum-connectivity Gourava index ($SGO$) demonstrates a highly competitive and low degeneracy value. For the Octane and Nonane datasets, $SGO$ yields degeneracy values of 22.22\% and 40.00\%, respectively, identical to the performance of the modern degree-based indicators ($SO$, $DSO$, $ABS$, and $N$).
    
    \item \textbf{Discrimination in Higher Orders:} In the case of all trees of order 10 ($N=106$), the $SGO$ index achieves a degeneracy of 42.45\%, matching the performance of the Sombor ($SO$) and diminished Sombor ($DSO$) indices. Notably, it marginally outperforms the atom-bond-sum connectivity ($ABS$) and Nirmala ($N$) indices, which both exhibit a degeneracy of 43.40\%.
\end{itemize}

In conclusion, the empirical results validate that the sum-connectivity Gourava index ($SGO$) possesses a significantly enhanced structural resolution compared to classical indices like $M_1$, $M_2$, and $H$, positioning it as a highly reliable tool for QSPR/QSAR studies requiring low structural degeneracy.

\subsection{On structure sensitivity}
The structural sensitivity of a topological index reflects its ability to respond consistently to changes in molecular structure. Ideally, a reliable topological descriptor should exhibit gradual variations when the underlying molecular structure is modified incrementally. Two measures have been studied in the literature to quantify this behavior \cite{13,14,15}. In this work, we employ the two parameters, namely, structure sensitivity ($SS$) and abruptness ($Abr$), to assess the performance of the considered topological indices. While $SS$ measures the response of an index to minor structural variations, $Abr$ identifies numerical abnormalities when the structure is only mildly altered. 

Let $G$ be a reference molecular graph, and let $S(G)$ denote the set of all molecular graphs obtained from $G$ through graph edit operations with graph edit distance (GED) equal to 2. Furthermore, let $TI(G)$ denote the value of a topological index for the graph $G$, $TI(H)$ denote the corresponding value for a graph $H \in S(G)$, and $|S(G)|$ represent the cardinality of the set $S(G)$. Using this notation, the structure sensitivity and abruptness of the topological index $TI$ are respectively defined by

\begin{equation*}
SS_G(TI)=
\frac{1}{|S(G)|}
\sum_{H\in S(G)}
\left|
\frac{TI(G)-TI(H)}{TI(G)}
\right|
\end{equation*}

\begin{equation*}
Abr_G(TI)=
\max_{H\in S(G)}
\left|
\frac{TI(G)-TI(H)}{TI(G)}
\right|
\end{equation*}

In addition, we consider the sensitivity--abruptness ratio ($SA$), recently introduced in \cite{16}, which is defined as $SA = SS/Abr$. The parameters $SS$, $Abr$, and $SA$ provide complementary information regarding the responsiveness, stability, and overall reliability of a topological index. These measures were computed for the selected indices over the complete set of all non-isomorphic trees of order $n=10$, and the corresponding results are summarized in Table~4.

\newpage
\begin{table}[hbt!]
\centering
\renewcommand{\arraystretch}{1.3}
\begin{tabular}{lrrr}
\toprule
\textbf{Index} & \textbf{Structure Sensitivity ($SS$)} & \textbf{Abruptness ($Abr$)} & \textbf{SA Ratio ($SA = SS/Abr$)} \\
\midrule
$M_1$ & 0.072624 & 0.162312 & 0.447438 \\
$M_2$ & 0.103238 & 0.270633 & 0.381578 \\
$H$   & 0.066918 & 0.138086 & 0.484672 \\
$S$   & 0.092601 & 0.207111 & 0.447105 \\
$DSO$& 0.019758 & 0.041318 & 0.478194 \\
$ABS$& 0.024709 & 0.052873 & 0.467338 \\
$SGO$& 0.034429 & 0.072575 & 0.474386 \\
$N$   & 0.035584 & 0.077521 & 0.459013 \\
\bottomrule
\end{tabular}
\caption{Structure Sensitivity ($SS$), Abruptness ($Abr$), and Structure-Abruptness Ratio ($SA$) across topological indices for non-isomorphic trees of order $n = 10$.}
\end{table}

Out of eight topological indices evaluated at $n = 10$, $SGO$ achieves the third-highest structure-abruptness ratio ($SA = 0.474386$), placing within $2.2\%$ of the top values ($H$ and $DSO$) and outperforming classical indices ($M_1$, $M_2$, $S$, and $N$) by $3.4\%$ to over $24\%$. Crucially, while $SGO$ trades away a negligible $0.8\%$ in $SA$ compared to $DSO$, it yields a $74\%$
higher structure sensitivity ($0.034429$ vs.\ $0.019758$). Thus, $SGO$ offers substantially greater structural discrimination at almost no cost in stability, making it a compelling candidate for QSAR/QSPR applications where telling similar molecules apart is critical.

\section{Conclusion}
In this study, we have demonstrated that both the sum-connectivity Gourava index and the product-connectivity Gourava index are closely related and effective molecular descriptors for characterizing the topological properties of molecules. Our comparative regression analysis reveals that the sum-connectivity Gourava index (as represented in (13)) provides a
significantly better fit for the relative contributions of various edge-types compared to the product-connectivity variant (as seen in (14)), with its coefficients closely aligning with the optimal least-squares results of (12).

Beyond regression fit, three further lines of evidence support the strength of $SGO$ as a topological descriptor. First, correlation analysis shows that while $SGO$ is strongly and negatively correlated with $M_1$, $M_2$, $SO$, $DSO$, $ABS$, and $N$ (coefficients ranging from $-0.8936$ to $-0.9923$), these values fall short of a perfect inverse relationship, indicating that
$SGO$ encodes structural information not fully captured by the other descriptors. Second, degeneracy analysis across the Octane, Nonane, and order-$10$ tree datasets shows that $SGO$ consistently achieves low degeneracy ($22.22\%$, $40.00\%$, and $42.45\%$, respectively), matching the best-performing modern indices ($SO$, $DSO$, $ABS$) and substantially
outperforming the classical $M_1$, $M_2$, and $H$ indices, confirming its strong capability to discriminate among non-isomorphic structures. Third, structure-sensitivity analysis on trees of order $n = 10$ shows that $SGO$ attains the third-highest structure-abruptness ratio ($SA = 0.474386$) among the eight indices examined, trailing $H$ and $DSO$ by less than $2.2\%$ while
offering $74\%$ greater structure sensitivity than $DSO$ at a negligible $0.8\%$ cost in $SA$.

Taken together, these findings suggest that the sum-connectivity Gourava index combines strong discriminative power, meaningful structural independence from established descriptors, and a favorable balance between sensitivity and numerical stability, offering superior predictive power for certain physicochemical properties relative to comparable indices. We feel this warrants further investigation into the mathematical properties and broader chemical applications of the sum-connectivity Gourava index in quantitative structure-property relationship (QSPR) studies.

 \section*{Funding} No funding is available for this study.
\section*{Author contributions}  HMN: Conceptualization, methodology, and original draft writing. UVCK: Methodology. APB, NN: Validation.
\section*{Data Availability}
The data used to support the work are cited within the text as references. 
\section*{Declarations}
\textbf{Ethical Approval} Not applicable.\\
\textbf{Conflict of interests} The authors declare that they have no known competing financial interests or personal relationships that could have appeared to influence the work reported in this paper.

\end{document}